\documentclass[pra,aps,showpacs,epsf,superscriptaddress,twocolumn]{revtex4}
\usepackage{color}
\usepackage{amsmath}
\usepackage{amssymb}
\usepackage{graphicx}
\usepackage{bm}

\def\be{\begin{equation}}
\def\ee{\end{equation}}

\begin{document}

\title{Counterflow in a doubly superfluid mixture of Bosons and Fermions}

\author{F. Chevy}
\affiliation{Laboratoire Kastler Brossel, ENS-PSL Research University, CNRS, UPMC-Sorbonne Universit\'es, Coll\`ege de France}

\begin{abstract}
In this article, we calculate the friction between two counter-flowing bosonic and fermionic superfluids. In the limit where the boson-boson and boson-fermion interactions can be treated within the mean-field approximation, we show that the force can be related to the dynamical structure factor of the fermionic component. Finally, we provide asymptotic expressions for weakly and strongly attractive fermions and show that the damping rate obeys simple scaling laws close to the critical velocity.
\end{abstract}


\maketitle

\section{Introduction}

%From the early days of superfluidity to the most recent experiments in ultra-cold atoms or exciton-polariton condensates,

The onset of frictionless flow in quantum fluids is probably one of the most intriguing macroscopic manifestations of quantum mechanics. Recent experiments on Bose-Fermi superfluid mixtures gave a new twist to this old question by probing the critical velocity of a superfluid counterflow \cite{Ferrier2014Mixture}. When two miscible superfluids flow through each-other, Castin {\em et al.} suggested a generalization of the celebrated Landau criterion where superfluidity is destroyed by the shedding of a pair of elementary excitations in the two systems \cite{castin2014landau}. Later-on, this scenario was supported by the study of the lifetime of the quasi-particles \cite{zheng2014quasiparticle,kinnunen2015induced} or by the calculation  of the hydrodynamic spectrum \cite{abad2014counter}.

The results put forward in \cite{castin2014landau} were based on heuristic arguments and were focusing on the velocity threshold above which the counter-flow is damped. In this work, we provide a full microscopic treatment of the friction in a superfluid counterflow and we determine the explicit velocity dependence of the damping force above the critical velocity. Assuming that the boson-boson and boson-fermion interactions can be treated within the mean-field approximation, we show that the force can be related to the dynamic structure factor of the fermionic superfluid. Although the general expression of the structure factor of an attractive Fermi gas is not known exactly in the crossover between the Bardeen-Cooper-Schrieffer (BCS) and molecular Bose-Einstein Condensate (BEC) regimes, we provide asymptotic expressions in the limits of weak and strong interactions where the fermionic component behaves respectively as an ideal Fermi gas and an hydrodynamic Bose-Einstein condensate of dimers \cite{zwerger2012BCSBEC}. We find that close to the critical velocity, the force obeys simple scaling laws, with exponent depending on the value of the fermion-fermion scattering length.

\section{General setting}

We consider a mixture between a fermionic superfluid and a weakly interacting Bose-Einstein condensate moving at velocities $\bm V_{\rm f}$ and $\bm V_{\rm b}$ respectively. The Hamiltonian of the system can be written as

\be
\widehat H=\widehat H_{\rm b}+\widehat H_{\rm f}+\widehat H_{\rm bf},
\ee
where $\widehat H_{\rm b,f}$ are the Hamiltonians of the bosons and the fermions and $\widehat H_{\rm bf}$ describes the coupling between the two species.
In the mean-field approximation, we have
 \be
 \widehat H_{\rm bf}=g_{\rm bf}\int d^3\bm r\sum_\sigma\widehat\varphi(\bm r)^\dagger\widehat\varphi(\bm r)\widehat\psi_\sigma^\dagger(\bm r)\widehat\psi_\sigma(\bm r),
 \ee
 where $\widehat \psi_\sigma$ is the field operator of spin $\sigma$ fermions, and $\widehat \varphi$ that of the bosons.
  Assuming that the BEC can also be described within the mean-field approximation, then the bosonic field operator can be expanded over the Bogoliubov modes
 \be
 \widehat \varphi(\bm r)=e^{i\bm k_b\cdot\bm r}\left[\sqrt{n_b}+\frac{1}{\sqrt{\Omega}}\sum_{\bm q}\left(u_{\bm q}\widehat b_{\bm q}e^{i\bm q\cdot\bm r}-v_{\bm q}\widehat b_{\bm q}^\dagger e^{-i\bm q\cdot\bm r}\right)\right],
 \ee
 where $\Omega$ is the quantization volume, $\hbar\bm k_b=m\bm V_b$, $n_b$ is the density of bosons and the $(u_{\bm q},v_{\bm q})$ are the usual Bogoliubov coefficients. The exponential prefactor describes the Galilean boost associated with the the motion of the condensate (see Appendix \ref{Appendix:Bogoliubov}). In addition, due to the motion of the BEC, the energies of the Bogoliubov modes are shifted by Doppler effect and their spectrum is given by

 \be
 E_{\bm q,b}=E^{(0)}_{\bm q,b}+\hbar\bm q\cdot\bm V_b,
 \label{Eq:Bogoliubov}
 \ee
 where $E^{(0)}_{\bm q,b} = \hbar c_b q\sqrt{1+q^2\xi_b^2/2}$ is the Bogoliubov spectrum (here $c_b$ is the sound velocity and $\xi_b$ is the healing length) \cite{stringaripitaevskii}.

 Expanding $\widehat H_{\rm bf}$ to first order in fluctuations of the bosonic field, we have

 \be
  \widehat H_{\rm bf}= \widehat H_{\rm bf}^{(0)}+ \widehat H_{\rm bf}^{(1)}+\widehat H_{\rm bf}^{(1)\dagger}+...
 \ee
with
\be
\widehat H_{\rm bf}^{(0)}=g_{\rm bf}n_{\rm b}\int d^3\bm r\sum_\sigma\widehat\psi_{\sigma}^\dagger\widehat\psi_\sigma(\bm r),
\ee
and
\be
\widehat H_{\rm bf}^{(1)}=g_{bf}\sqrt{\frac{n_b}{\Omega}}\int d^3\bm r\sum_{\bm q,\sigma} (u_{\bm q}-v_{\bm q})e^{-i\bm q\cdot\bm r}\widehat\psi_{\sigma}^\dagger(\bm r)\widehat\psi_\sigma(\bm r)\widehat b_{\bm q}^\dagger.
\ee

$\widehat H_{\rm bf}^{(0)}$ being proportional to the number of fermions, it is a constant of motion, and can therefore be gauged away. $\widehat H_{\rm bf}^{(1)}$ and $\widehat H_{\rm bf}^{(1)\dagger}$ respectively create and destroy Bogoliubov excitations in the BEC. In the weak coupling limit we can expand the damping rate in power of $g_{\rm bf}$. To first order, only $\widehat H_{\rm bf}^{(1)}$ plays a role and using Fermi's Golden Rule the rate of creation of excitations of momentum $\hbar\bm q$ in the BEC and in an eigenstate $|\alpha_f\rangle$ of the fermionic Hamiltonian $\widehat H_{\rm f}$ is given by

\be
\begin{split}
\Gamma(\alpha,\bm q)=&\frac{g_{\rm bf}^2n_b}{\Omega h}(u_{\bm q}-v_{\bm q})^2\delta(E_{\alpha,f}+E_{\bm q,b})\times \\
&\left|\langle\alpha_f|\int d^3\bm r\sum_\sigma \widehat\psi_{\sigma}^\dagger\widehat\psi_\sigma(\bm r)e^{-i\bm q\cdot\bm r}|0_f\rangle\right|^2,
\end{split}
\ee
where $E_{\alpha,f}$ is the energy of the state  $|\alpha_f\rangle$ relatively to that  of  the ground state $|0_f\rangle$  of the moving fermionic superfluid. Since the creation of a Bogoliubov excitation imparts a momentum $\hbar\bm q$ to the BEC, the force acting on the bosons can be written as
\begin{eqnarray}
\bm F&=&\sum_{\alpha,\bm q}\hbar\bm q\Gamma(\alpha,\bm q)\\
&=&2\pi\frac{g_{\rm bf}^2n_b}{\Omega}\sum_{\bm q}\bm q(u_{\bm q}-v_{\bm q})^2S(-\bm q,-E_{\bm q,b}),
\label{EqForce}
\end{eqnarray}
where we introduced the dynamical structure factor of the Fermi gas
\be
S(\bm q,E)=\sum_{\alpha}\delta(E-E_{\alpha,f})\left|\langle\alpha_f|\int d^3\bm r\sum_\sigma \widehat\psi_{\sigma}^\dagger\widehat\psi_\sigma(\bm r)e^{i\bm q\cdot\bm r}|0_f\rangle\right|^2.
\ee
Note that the expression for the force is galilean invariant and depends only on the relative velocity $\bm V_{\rm b}-\bm V_{\rm f}$. Indeed, one can show that the dynamical structure factor $S$ of a system moving at velocity $\bm V_f$ is the related to the one of a system at rest (denoted as $S_{\rm st}$) by

\be
S(\bm q,E)=S_{\rm st}(\bm q,E-\hbar\bm q\cdot\bm V_f)
\ee
(this result is a direct consequence of Eq. (\ref{Eq:Energy}) giving the energy in the moving frame, and the fact that the operator $\int d^3\bm r\sum_\sigma \widehat\psi_{\sigma}^\dagger\widehat\psi_\sigma(\bm r)e^{i\bm q\cdot\bm r}$ gives a kick $\hbar\bm q$ to the system).  Inserting this expression in Eq. (\ref{EqForce}) and using the expression for the Bogoliubov spectrum of a moving BEC, Eq. (\ref{Eq:Bogoliubov}) we have for the force

\be
\bm F=2\pi\frac{g_{\rm bf}^2n_b}{\Omega}\sum_{\bm q}\bm q(u_{\bm q}-v_{\bm q})^2S_{\rm st}(-\bm q,-E^{(0)}_{\bm q,b}-\hbar\bm q\cdot\bm V),
\ee
with $\bm V=\bm V_f-\bm V_b$. Finally, using the explicit form of the Bogoliubov coefficients, we have
\be
(u_{\bm q}-v_{\bm q})^2=\frac{\varepsilon_{\bm q,b}}{E^{(0)}_{\bm q,b}}
\ee
with $\varepsilon_{\bm q,b}=\hbar^2 q^2/2m_b$, hence
\be
\bm F=2\pi\frac{g_{\rm bf}^2n_b}{\Omega}\sum_{\bm q}\bm q\frac{\varepsilon_{\bm q,b}}{E^{(0)}_{\bm q,b}}S_{\rm st}(-\bm q,-E^{(0)}_{\bm q,b}-\hbar\bm q\cdot\bm V).
\label{EqForceFinal}
\ee

Eq. (\ref{EqForceFinal}) is the main result of this paper and shows that the  friction  between the two superfluids is directly related to the dynamic structure factor of the fermionic component of the system. The dynamical structure factor describes the response of a given system to a sinusoidal perturbation and can be related to two-body correlations. In strongly correlated gases,  it was measured using Bragg spectroscopy  \cite{veeravalli2008bragg} and was used to determine the value of Tan's contact parameter describing the tail of the momentum distribution \cite{Kuhnle2010universal}.

\begin{figure}
\centerline{\includegraphics[width=\columnwidth]{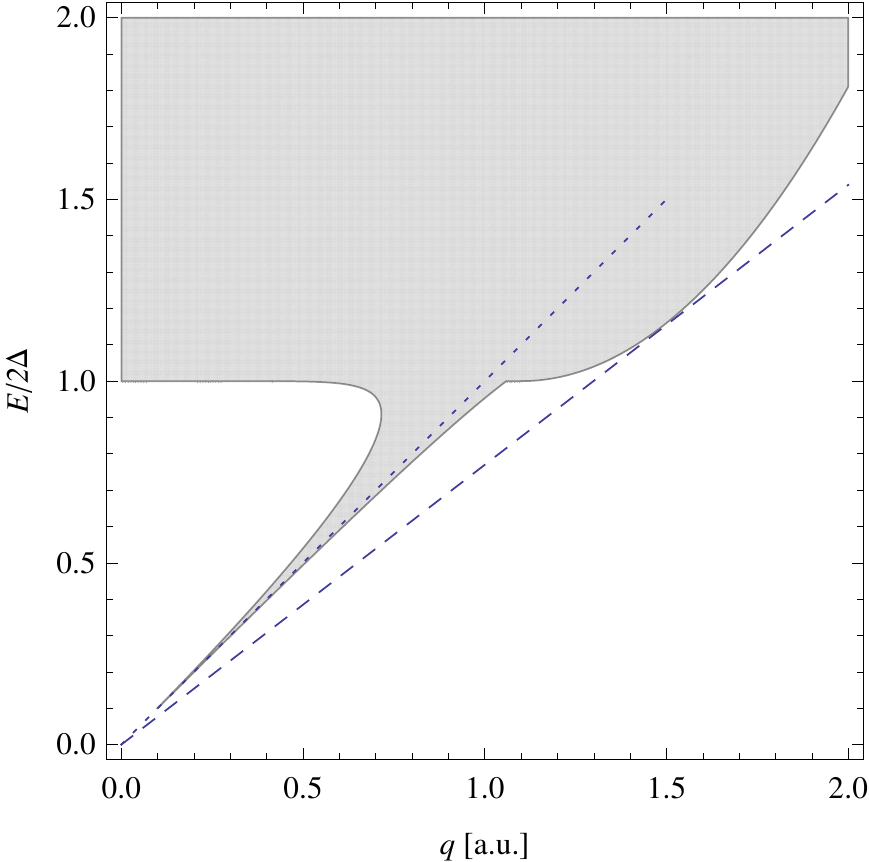}}
\caption{Sketch of the structure factor of an interacting Fermi gas: $S_{\rm st}$ vanishes outside the gray area and $\Delta$ is the excitation gap. Dotted line: low momentum phonon branch $E=\hbar c_{\rm f} q$. Dashed line: Landau's critical velocity $E=\hbar v_{\rm L}q$ below which the structure factor vanishes. Note that this sketch corresponds to a weakly interacting system where Landau's velocity is fixed by pair breaking, and $v_{\rm L}<c_{\rm f}$. In the molecular regime of the crossover, $v_L=c_{\rm f}$ \cite{Combescot2006,castin2014landau}.}
\label{Fig1}
\end{figure}

The properties of strongly correlated Fermi systems are notoriously hard to calculate, and approximate expression for the dynamic structure factor have been obtained using the approximate methods  \cite{minguzzi2001dynamic,buchler2004spectroscopy}. However, its general properties can be summarized on the sketch of Fig. \ref{Fig1}. At low momenta, the excitation spectrum is dominated by a phonon branch $E=\hbar c_{\rm f}k$, with $c_{\rm f}$ the sound velocity of the fermionic superfluid. Since the ground state defines the energy zero, $S_{\rm st}$ vanishes for $E< 0$. However, the presence of a pairing gap implies of a region where $S_{\rm st}$ vanishes at positive energies. Landau's critical velocity $v_{\rm L}$ is defined as the highest value of $v$ such that $S_{\rm st}(E,q)$ vanishes for all $E\le v q$ (see Fig. \ref{Fig1}). In the BEC regime of the BEC-BCS crossover, we have $v_{\rm L}=c_{\rm F}$, while in the BCS regime, Landau's velocity is driven by pair breaking excitations and we have $v_{\rm L}\ll c_{\rm f}$.

 These general features implies that if for all $\bm q$ we have $-E^{(0)}_{\bm q,b}-\hbar\bm q\cdot\bm V< qv_{\rm L}$, the force is zero. To get some damping, there must be values of $\bm q$ for which  $-E^{(0)}_{\bm q,b}-\hbar\bm q\cdot\bm V\ge \hbar q v_{\rm L}$. Using the general Landau argument, we get that
$|\bm V|\ge v_{\rm L}+\min_q (E^{(0)}_{\bm q,b}/\hbar q)=v_{\rm L}+c_{\rm b}$. Note that this bound sets a necessary, but not  sufficient, condition on the relative velocity. As a consequence $v_{\rm L}+c_{\rm b}$ is only a lower bound for the real critical velocity.

Another consequence of this inequality is that the values of $q$ contributing to the integral are bounded. Indeed, from the  inequality  $E\le \hbar  q v_L$, we get that

$$
q\xi_{\rm b}\le \sqrt{2 x(x+2)},
$$
where $xc_{\rm b}=V-c_{\rm b}-v_{\rm L}$. In other words, when the relative velocity gets close to $c_{\rm b}+v_{\rm L}$, the damping is dominated by long wave-length excitations.

In what follows, we compute the force in the deep BEC and BCS limits where analytical expressions can be obtained.

\section{BCS limit}
Let's consider first the deep BCS limit where the fermions can be described as an ideal gas. Since there is a vanishingly small  gap in the fermionic spectrum the lower bound derived above is actually saturated and we have $V_c=c_b$. Working close to $c_b$, the relevant values of $q$ much smaller than $k_F$, and we can therefore replace the expression of the dynamical structure factor by its low-energy approximation (see for instance \cite{baerwinkel1971dynamic}, and Appendix \ref{AppendixStructureFactor})

$$
S_{\rm st}(\bm q,E)=\frac{m_{\rm f}^2\Omega E}{2\pi^2\hbar^4 q},
$$
this replacement being valid as long as the condition
$$
k_F\xi\ll \sqrt{2\left(\frac{V^2-c_{\rm b}^2}{c_{\rm b}^2}\right)}
$$
is fulfilled.

Replacing the discrete sum over $q$ by an integral, one gets

 \be
 \frac{\bm F}{\Omega}=-\frac{2 g_{\rm bf}^2 n_{\rm b}m_{\rm f}^2}{(2\pi)^4\hbar^4}\int d^3\bm q \frac{\bm q\varepsilon_{{\rm b},\bm q}}{qE^{(0)}_{{\rm b},\bm q}}(E_{{\rm b},\bm q}^{(0)}+\hbar \bm q\cdot\bm V)\Theta(-E_{{\rm b},\bm q}^{(0)}-\hbar \bm q\cdot\bm V).
 \ee
where $\Theta$ is Heaviside's step function. The integral can be computed analytically and taking $V'=V/c_{\rm b}$, the force becomes
 \be
 \frac{\bm F}{\Omega}=-\frac{g_{\rm bf}^2 n_{\rm b}m_{\rm f}^2}{(2\pi)^3m_{\rm b}\hbar^2\xi^5}I(V'),
 \ee
with
\be
\begin{split}
I(V')=&\frac{1}{105 \sqrt{2} V'^2}\left[105 V'^3 \sinh ^{-1}\left(\sqrt{V'^2-1}\right)+\right.\\
&\left.\sqrt{V'^2-1} \left(6 V'^6-39 V'^4-80
   V'^2+8\right)\right].
\end{split}
\ee
Close to $V'=1$, this expression can be expanded as
\be
I(V')\simeq \frac{128}{315} \left(V'-1\right)^{9/2}+...
\ee

Even though this scaling is strictly-speaking obtained for a non-interacting Fermi gas, it can be extended to the case of a BEC moving inside an interacting Fermi gas in its normal phase when it can be described within the Fermi Liquid Theory. Indeed, close the critical velocity, the damping mechanism described involves only low energy excitations of the system which behave as fermionic quasiparticles with renormalized physical parameters. In particular, the scaling $S(q,E)\propto E/q$ is still valid \cite{stringari2009density}.

\section{Hydrodynamic Limit}

In the BEC limit of the cross-over, the excitations of the fermionic superfluid are dominated by phonons of energy $E_{f,\bm q}$. In this case, the dynamical structure factor can be approximated by a Dirac function and we have in this case

\be
S_{\rm st}(\bm q,E)={\cal A}\delta (E-E_{f,\bm q}^{(0)})
\label{HydroS}
\ee
where the normalization constant $\cal A$ is fixed by the f-sum rule \cite{pines1966nozieres,stringaripitaevskii} and is given by
\be
{\cal A}=N_f\frac{\varepsilon_{q,f}}{E_{f,\bm q}^{(0)}}.
\ee
In this limit, we obtain for the force
\be
\bm F=2\pi g_{\rm bf}^2n_b n_f\sum_{\bm q}\bm q\frac{\varepsilon_{\bm q,b}\varepsilon_{\bm q,f}}{E_{\bm q,b}^{(0)}E_{-\bm q,f}^{(0)}}\delta(E^{(0)}_{\bm q,b}+E^{(0)}_{-\bm q,f}+\hbar\bm q\cdot\bm V).
\label{EqForceHydro}
\ee
Using the usual Landau argument, we see that the force vanishes if the velocity is smaller than a critical velocity $V_c$ given by
\be
V_c=\min_{\bm q}\left(\frac{E^{(0)}_{\bm q,b}+E^{(0)}_{-\bm q,f}}{\hbar q}\right).
\ee
In case of linear dispersion relation, the critical velocity is simply
\be
V_c=c_{\rm b}+c_{\rm f},
\ee
where $c_{\rm f,b}$ are the sound velocities of the fermions and the bosons.

To calculate explicitly the force, we assume that both the bosons and the fermions follow the Bogoliubov dispersion relation, an assumption valid in the far-BEC limit of the BEC-BCS crossover. Taking $\xi_{\rm b,f}$ the healing lengths of the two superfluids, we have after performing the angular integral

\be
\begin{split}
\frac{F}{\Omega}=&-\frac{g_{\rm bf}^2 n_{\rm b}n_{\rm f}\hbar}{8\pi m_{\rm f}m_{\rm b}c_{\rm b}c_{\rm f}V}\times\\
&\int_0^{q_M} \frac{q^4dq\left(\frac{c_{\rm b}}{V}\sqrt{1+q^2\xi_b^2/2}+\frac{c_{\rm f}}{V}\sqrt{1+q^2\xi_f^2/2}\right)}{\sqrt{(1+q^2\xi_b^2/2)(1+q^2\xi_f^2/2)}}
\end{split}
\ee
with $q_M$ given by the condition
\be
c_{\rm b}\sqrt{1+q_M^2\xi_{\rm b}^2/2}+c_{\rm f}\sqrt{1+q_M^2\xi_{\rm f}^2/2}=V
\ee

For $V\simeq V_c$, $q_M$ is vanishingly small, and is given by

$$
q_M\simeq 2\sqrt{\frac{V-V_c}{c_{\rm b}\xi_b^2+c_{\rm f}\xi_f^2}},
$$
hence
\be\frac{F}{\Omega}\simeq-\frac{4g_{\rm bf}^2 n_{\rm b}n_{\rm f}\hbar}{5\pi m_{\rm f}m_{\rm b}c_{\rm b}c_{\rm f}V_c}\left(\frac{V-V_c}{c_{\rm b}\xi_b^2+c_{\rm f}\xi_f^2}\right)^{5/2}
\ee

\section{Conclusion}
In this work we have shown that the damping between two bosonic and fermionic superfluids could be related to the dynamical structure factor of the fermions. In the weakly and strongly attractive limits, we find a power law dependence of the damping force vs velocity close to the critical velocity. Like previous findings \cite{zheng2014quasiparticle}, we find that the exponent varies across the BEC-BCS crossover and is different from the scaling $F\propto (V'-1)^2$ found for the drag of a particle in a Bose-Einstein condensate \cite{astrakharchik2004motion}. These different scalings highlight the many-body nature of the two counter-flowing systems and the role of the collective excitations in the damping mechanism. Our prediction can be directly tested by measuring the damping of the relative oscillations of two superfluids, as presented in \cite{Ferrier2014Mixture} for a unitary fermions. In particular, comparison with Eq. \ref{EqForceFinal} provides a quantitative test of theoretical predictions for the structure factor of a strongly correlated fermionic superfluid.

\begin{acknowledgments}
The authors thank X. Leyronas, C. Salomon, Y. Castin, I. Ferrier-Barbut, S. Laurent, M. Delehaye and S. Stringari for stimulating discussions. This work was supported by R\'egion \^Ile de France (IFRAF) and European Union (ERC grant ThermoDynaMix).
\end{acknowledgments}

\appendix

\section{Bogoliubov Transformation in a Moving Frame}
\label{Appendix:Bogoliubov}
Consider a ensemble of $N$ particles described by a state $|\psi\rangle$. The action of a velocity $\bm V$  boost  is described by the operator
\be
\widehat U=\exp\left(i\sum_{j=1}^N\bm k\cdot\widehat{\bm r_i}\right),
\ee
with $\bm k=m\bm V/\hbar$.
In second quantization, this operator is represented by
\be
\widehat U=\exp\left(i\int d^3\bm r'\widehat\varphi(\bm r')^\dagger\widehat\varphi(\bm r')\bm k\cdot\bm r'\right).
\ee

Define $\widehat\varphi'(\bm r)$ the field operator in the moving frame. By definition, we must have for $|\psi'_{i=1,2}\rangle=\widehat U|\psi_{i=1,2}\rangle$, $\langle\psi'_1|\widehat\varphi'(\bm r)|\psi'_2\rangle=\langle\psi_1|\widehat\varphi(\bm r)|\psi_2\rangle$, hence
\be
\langle\psi_1|\widehat U^\dagger\widehat\varphi'(\bm r)\widehat U|\psi_2\rangle=\langle\psi_1|\widehat\varphi(\bm r)|\psi_2\rangle.
\ee
This equation being valid for any $|\psi_i\rangle$, we must have $\widehat U^\dagger\widehat\varphi'(\bm r)\widehat U=\widehat\varphi(\bm r)$ hence
\be
\widehat\varphi'(\bm r)=\widehat U\widehat\varphi(\bm r)\widehat U^\dagger.
\ee
$\widehat\varphi'$ can be calculated using the Campbell-Hausdorff formula. Taking $\widehat U=\exp(\widehat T)$, we have indeed
\be
\widehat\varphi'(\bm r)=\widehat\varphi(\bm r)+[\widehat T,\widehat\varphi(\bm r)]+\frac{1}{2}[\widehat T,[\widehat T,\varphi(\bm r)]]+...
\ee
The commutator $[\widehat T,\widehat\varphi(\bm r)]$ is readily calculated and yields $[\widehat T,\widehat\varphi(\bm r)]=-i\bm k\cdot\bm r\widehat\varphi(\bm r)$. After resumming the Campbell-Hausdorff formula, we obtain
\be
\widehat\varphi'(\bm r)=e^{-i\bm k\cdot\bm r}\widehat\varphi(\bm r).
\label{Eq:galilean}
\ee
In the moving frame, the BEC is stationary and we can therefore decompose $\varphi'(\bm r)$ using a Bogoliubov transform
\be
\widehat\varphi'(\bm r)=\sqrt{n_0}+\frac{1}{\sqrt{\Omega}}\sum_{\bm q}\left[u_{\bm q}e^{i\bm q\cdot\bm r}\widehat b_{\bm q}-v_{\bm q}e^{-i\bm q\cdot\bm r}\widehat b_{\bm q}^\dagger\right].
\ee
We thus have in the lab frame
\begin{eqnarray}
\widehat\varphi(\bm r)&=&e^{i\bm k\cdot\bm r}\widehat\varphi'(\bm r)\\
&=&e^{i\bm k\cdot\bm r}\left[\sqrt{n_0}+\frac{1}{\sqrt{\Omega}}\sum_{\bm q}\left(u_{\bm q}e^{i\bm q\cdot\bm r}\widehat b_{\bm q}-v_{\bm q}e^{-i\bm q\cdot\bm r}\widehat b_{\bm q}^\dagger\right)\right].
\end{eqnarray}
The eigenenergies of the Bogoliubov excitations of the moving Bose-Einstein condensate are obtained by applying the Galilean boost operator $\widehat U$ to the eigenstates $|\beta\rangle_{\rm b}$ of the condensate at rest. Taking $|\beta'\rangle_{\rm b}=\widehat U|\beta\rangle_{\rm b}$, and consider the quantum many-body Hamiltonian
\be
\begin{split}
\widehat H_{\rm b}=&-\int \widehat\varphi(\bm r)^\dagger\frac{\hbar^2\nabla^2}{2m}\widehat\varphi(\bm r)d^3\bm r\\
&+\int  V(\bm r-\bm r')\widehat\varphi(\bm r)^\dagger\widehat\varphi(\bm r')^\dagger \widehat\varphi(\bm r')\widehat\varphi(\bm r)^\dagger d^3\bm r d^3\bm r'.
\end{split}
\ee
By using Eq. \ref{Eq:galilean}), we see readily that if $|\beta\rangle_{\rm b}$ is a common eigenstate of $\widehat H_{\rm b}$ and the momentum and particle number operators, then $|\beta'\rangle_{\rm b}$ is an eigenstate of $\widehat H_{\rm b}$ for the eigenvalue
\be
E'_\beta=E_\beta+\hbar\bm q_\beta\cdot\bm V+N_{\rm b}\frac{mV^2}{2},
\ee
where $\hbar \bm q_\beta$ is the momentum of state $|\beta\rangle_{\rm b}$. Compared to the energy $E'_0$ of the boosted ground-state Bose-Einstein condensate, the energy of the excitation $|\beta'\rangle_{\rm b}$ is therefore
\be
E'_\beta-E'_0=E_\beta+\hbar\bm q_\beta\cdot\bm V,
\label{Eq:Energy}\ee
and is simply Doppler-shifted.

\section{Dynamic structure factor of the ideal Fermi gas.}
\label{AppendixStructureFactor}

\begin{figure}
\centerline{\includegraphics[width=0.85\columnwidth]{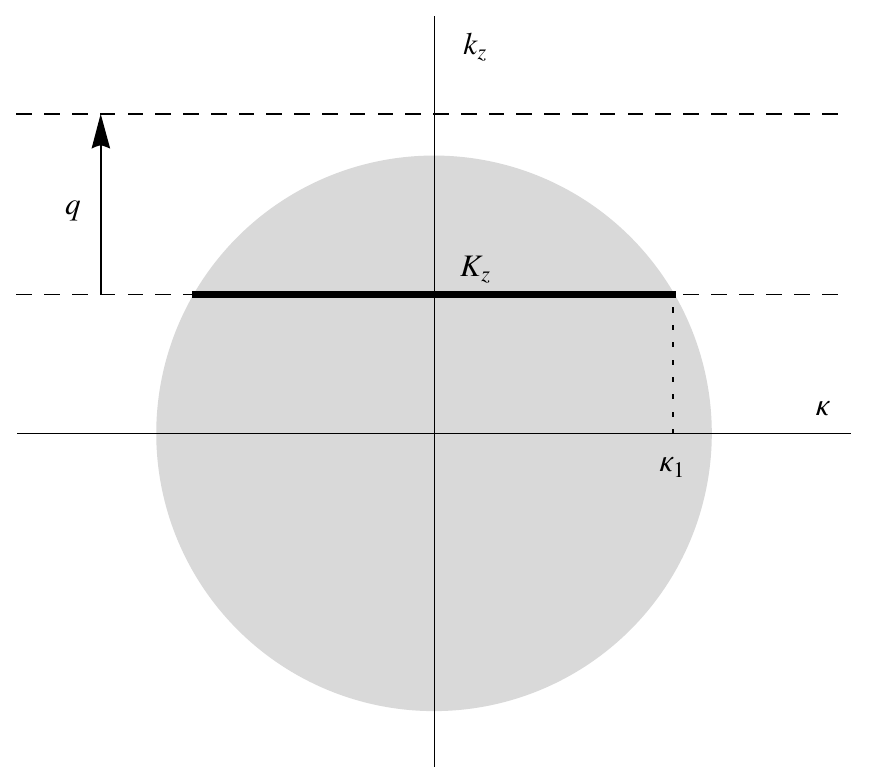}}
\centerline{\includegraphics[width=0.85\columnwidth]{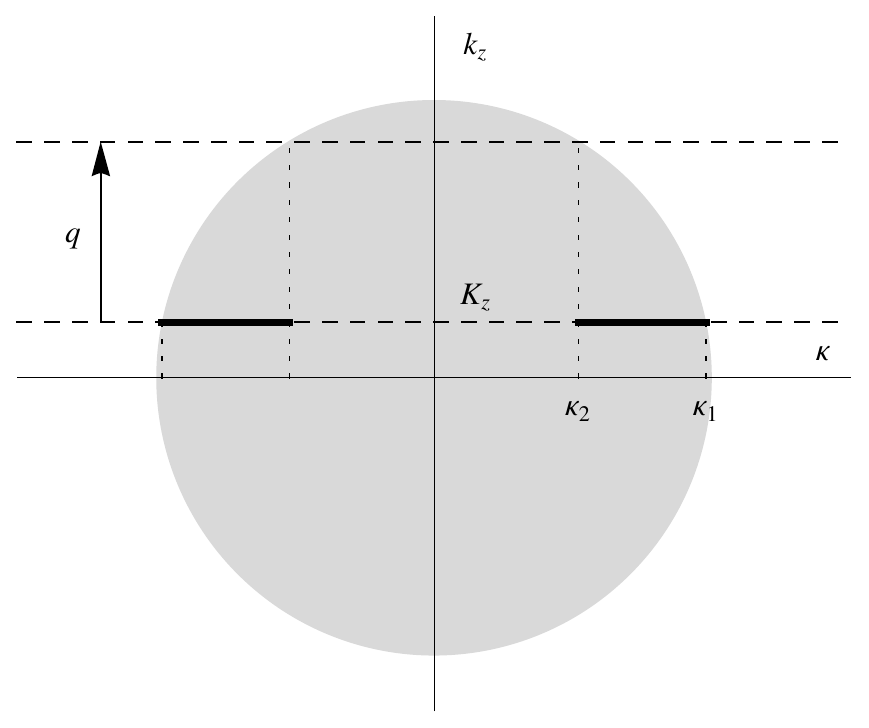}}
\centerline{\includegraphics[width=0.85\columnwidth]{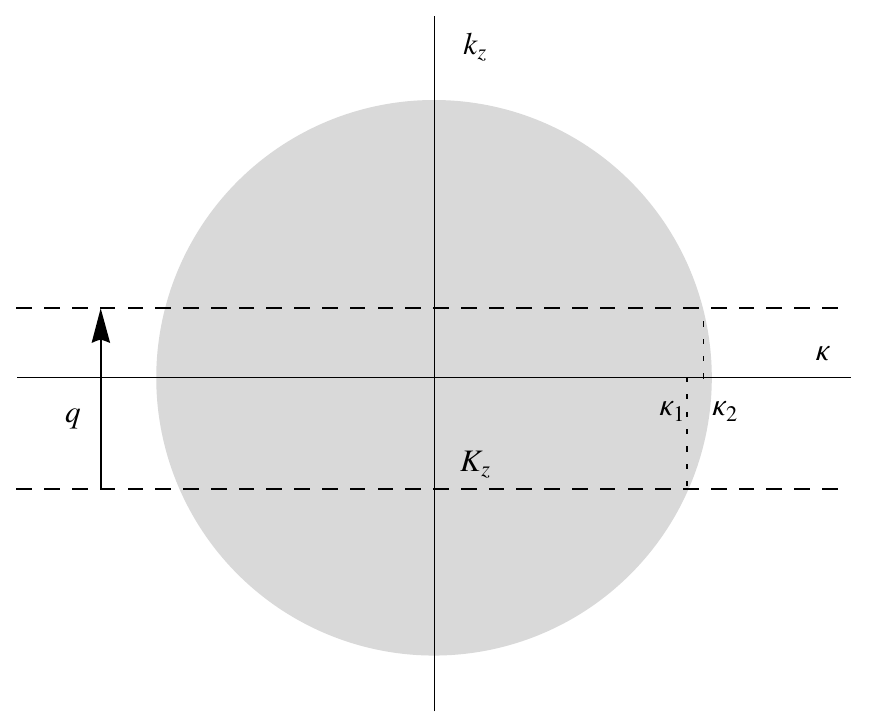}}
\caption{Construction of the dynamic structure factor in momentum space. The shaded area represents the Fermi sphere. Energy-momentum conservation fixes the initial $k_z$ to a value $K_z$ (see text). Since the initial state must lie inside the Fermi Sea, we must have $\kappa\le \kappa_1=\sqrt{k_F^2-K_z^2}$. Moreover, the final momentum must be above the Fermi surface, hence $\kappa\ge \sqrt{k_F^2+(K_z+q)^2}$. The thick solid line represents the momenta contributing to the dynamical structure factor. From top to bottom. Upper panel, case (1): $K_z+q$ larger than $k_f$. The final momentum is always above the Fermi surface. Middle panel, case 2: $K_z+q$ is lower than $k_F$, with $\kappa_2\le \kappa_1$. The permitted values of $\bm k$ are located in the ring $\kappa_2\le \kappa\le\kappa_1$. Lower panel, case 3. For $\kappa_2\ge \kappa_1$, the contribution to the structure factor vanishes.  }
\label{FigFermi}
\end{figure}

 In this appendix, we rederive the dynamical structure factor for zero-temperature, non-interacting fermions. A more general derivation including finite temperature effects can be found in \cite{baerwinkel1971dynamic}.

The dynamic structure factor is given by the matrix elements of the excitation operator $\widehat F_{\bm q}=\sum_\sigma\int d^3\bm r e^{i\bm q\cdot\bm e}\widehat\psi_\sigma(\bm r)^\dagger\widehat\psi_\sigma(\bm r)$. Expanding the field operators over the plane wave basis as $\widehat \psi_\sigma(\bm r)=\sum_{\bm k}e^{i\bm k\cdot\bm r}\widehat c_{\bm k,\sigma}/\sqrt{\Omega}$, where $\Omega$ is a quantization volume, yields

$$
\widehat F_{\bm q}=\sum_{\bm k,\sigma}\widehat c_{\bm k+\bm q,\sigma}^\dagger\widehat c_{\bm k,\sigma}.
$$
In other words, the operator $\widehat F_{\bm q}$ kicks one particles of the system and gives it a momentum $\bm q$. The states $|n\rangle$ excited by $\widehat F_{\bm q}$ from the ground state $|0\rangle$ correspond to particle-hole pairs of momenta $\bm k$ (hole) and $\bm k+\bm q$ (particle). The energy of such an excitation is $E_n=\varepsilon_{\bm k+\bm q}-\varepsilon_{\bm k}=\hbar^2q^2/2m+\hbar^2\bm k\cdot\bm q/m$ and we can write the structure factor as

$$S(\bm q,E)=2\int_{k<k_F,|\bm k+\bm q|>k_F} \frac{d^3\bm k\Omega}{(2\pi)^3}\delta\left(E-\frac{\hbar^2q^2}{2m}-\frac{\hbar^2k_z q}{m}\right),$$
where the $z$ axis is chosen along $\bm q$ and where the factor 2 comes from the two spin states.

We perform first the integral over $k_z$. Take $K_z(\bm q,E)=(E-\frac{\hbar^2q^2}{2m})m/\hbar^2q$. The delta-function selects $k_z=K_z(\bm q,E)$ and taking $\bm\kappa=(k_x,k_y)$ we have

\be
S(\bm q,E)=\frac{2m}{\hbar^2q}\int_{|K_z|<k_F,|\bm k+\bm q|>k_F} \frac{d^2\bm \kappa\Omega}{(2\pi)^3}.
\label{EqIntKappa}
\ee

The condition $|K_z|\le k_F$ selects a band in the $(q,E)$ plane. Indeed, we must have

$$
-\frac{\hbar^2 k_F q}{m}\le \left(E-\frac{\hbar^2 q^2}{2m}\right)\le \frac{\hbar^2 k_F q}{m}.
$$

The constraint $k_z=K_z$ selects a plane in initial-momentum space. Taking into account the fact that the initial momenta must lie within the Fermi Sea, we see that the transverse momentum $\bm\kappa$ lies within a disk of radius $\kappa_1=\sqrt{k_F^2-K_z^2}$. The set of allowed values for $\bm\kappa$ is further constrained by Pauli Blocking on the final states. Indeed, the final momentum must lie outside the Fermi Sea and as such it implies the condition $|\bm k+\bm q|\ge k_F$. We must then consider three cases (see also Fig. \ref{FigFermi}) :

\begin{enumerate}
\item If $K_z+q$ is larger than $k_F$, then for any  $\bm k$ such that $k_z=K_z$, the final momentum $\bm k+\bm q$ is outside the Fermi Sea. The constraint on the final state is thus always satisfied and does not have any effect on the calculation of the integral over $\bm\kappa$, which yields the area of the disk of radius  $\kappa_1$. We thus have

    \be
    S(\bm q,E)=\frac{m\Omega \kappa_1^2}{4\pi^2\hbar^2 q}.
    \ee
\item For $K_z+q\le k_F$, final momenta corresponding to small values of $\kappa$ (more precisely smaller than $\kappa_2=\sqrt{k_F^2-(K_z+q)^2}$) lie inside the Fermi Sea and are therefore forbidden by Pauli Principle. We have then

    \be
    S(\bm q,E)=\frac{m\Omega (\kappa_1^2-\kappa_2^2)}{4\pi^2\hbar^2 q}=\frac{m^2\Omega E}{2\pi^2\hbar^4 q}.
    \label{EqS2}
    \ee

 Take $k_F^3=3\pi^2 n_{\rm f}$, where $n_{\rm f}$ is the total density of fermions, this expression can be recast as

 \be
 S(\bm q,E)=\frac{3N_{\rm f}}{8}\left(\frac{k_F}{q}\right)\frac{E}{E_F^2},
\ee
with $N_{\rm f}=n_{\rm f}\Omega$ the total number of fermions.

\item Eq. \ref{EqS2} is valid as long as $\kappa_1\ge \kappa_2$. Graphically, case (2) ends when $K_z+q=-K_z$, ie for $E=0$. For $E\le 0$, the dynamic structure factor vanishes, as expected since the Fermi Sea is the ground state of the system.
\end{enumerate}

\begin{figure}
\centerline{\includegraphics[width=\columnwidth]{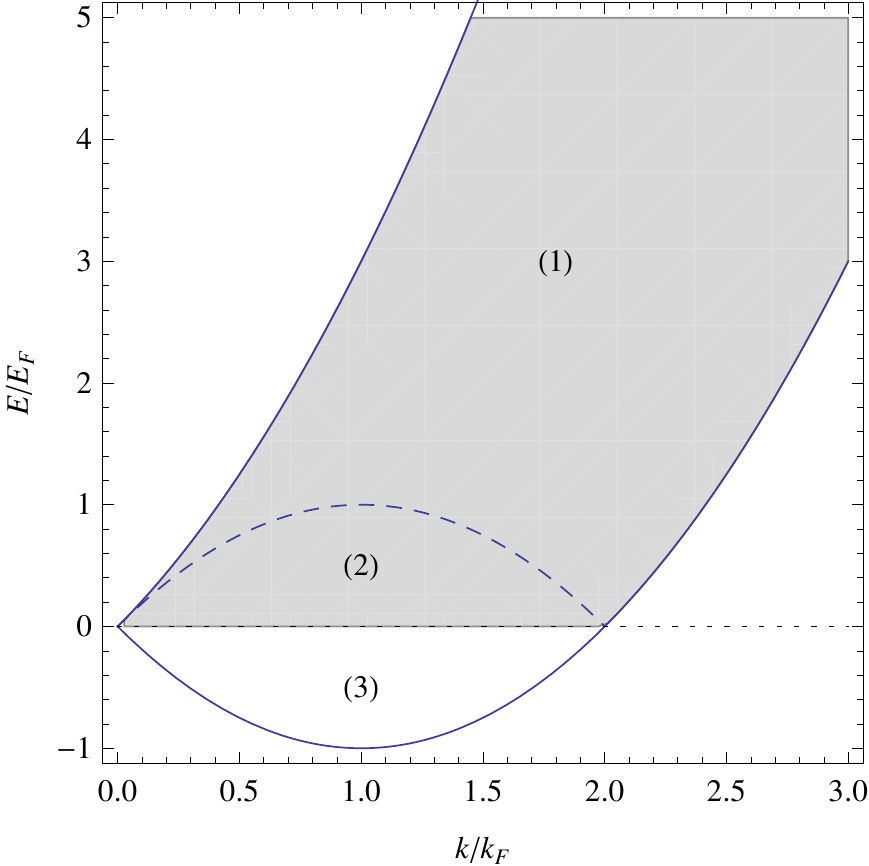}}
\caption{Dynamical structure factor of the ideal gas. The shaded area corresponds to the region where the structure factor takes finite values. The solid lines corresponds to condition $-k_F\le K_z\le k_F$. The dashed marks the limit between cases (1) and (2) and is associated with the condition $K_z+q=k_F$.  }
\label{Fig:S}
\end{figure}

 Finally, let's consider some simple limiting cases. For $q\ge 2 k_F$,  we are in case (1) (see Fig. \ref{Fig:S}), and $S(\bm q,E)$ is therefore given by

\be
S(\bm q,E)=\frac{m\Omega}{4\pi^2\hbar^2 q}\left[k_F^2-\frac{m^2}{\hbar^2q^2}\left(E-\frac{\hbar^2 q^2}{2m}\right)^2\right].
\label{EqSHigh}
\ee

In this regime, $S$ is peaked around the energy $E=\varepsilon_{\bm q}=\hbar^2 q^2/2m$ with a half-width $\Delta E=\hbar^2 k_F q/m$. We note that $\Delta E/\varepsilon_{\bm q}=2k_F/q\rightarrow 0$ for $q\rightarrow\infty$. In other words, we recover at large momenta the dynamic structure factor of a free particle.

For low values of $q$, the range of validity of case (1) is vanishingly small and the dynamic structure factor is mostly given by case (2). We therefore have
 \be
 S(\bm q,E)=\frac{3N_{\rm f}}{8}\left(\frac{k_F}{q}\right)\frac{E}{E_F^2},
\label{EqSLow}
\ee
as long as $0\le E\le 2E_F q/k_F$, and $S=0$  otherwise.

One readily checks that both approximate expressions (\ref{EqSLow}) and (\ref{EqSHigh}) satisfy the f-sum rule.

\bibliographystyle{unsrt}
\bibliography{bibliographie}

 \end{document}